\documentclass[12pt]{article}
\usepackage[reqno]{amsmath}
\usepackage{bbm}
\usepackage{epsfig}
\usepackage{array}
\usepackage{float}

\usepackage{a4}

\usepackage{a4wide}
\usepackage{wasysym}

\def\ra{\rightarrow}
\def\be{\begin{equation}}
\def\ee{\end{equation}}
\def\gs{\mathrel{
   \rlap{\raise 0.511ex \hbox{$>$}}{\lower 0.511ex \hbox{$\sim$}}}}
\def\ls{\mathrel{
   \rlap{\raise 0.511ex \hbox{$<$}}{\lower 0.511ex \hbox{$\sim$}}}}

\newcommand{\onbb}{neutrinoless double beta decay }
\newcommand{\ba}{\begin{array}{c}}
\newcommand{\baz}{\begin{array}{cc}}
\newcommand{\bad}{\begin{array}{ccc}}
\newcommand{\bea}{\begin{equation} \begin{array}{c}}
\newcommand{\eea}{ \end{array} \end{equation}}
\newcommand{\ea}{\end{array}}
\newcommand{\D}{\displaystyle}
\newcommand{\dms}{\mbox{$\Delta m^2_{\odot}$}}
\newcommand{\dma}{\mbox{$\Delta m^2_{\rm A}$}}
\newcommand{\meff}{\mbox{$\langle m \rangle$}}

\begin{document}

\title{\vspace{-2cm}
\hfill {\small OUTP-0422P}\\
\vspace{-0.3cm}
\hfill {\small TUM-HEP-566/04}\\
\vspace{-0.3cm}
\hfill {\small SISSA 81/2004/EP}\\
\vspace{-0.3cm}
\hfill {\small hep-ph/0411190} 
\vskip 0.2cm
\bf A Flavor Symmetry for quasi-degenerate Neutrinos: $L_\mu - L_\tau$
}
\author{
Sandhya Choubey$^a$\thanks{email: \tt sandhya@thphys.ox.ac.uk}\mbox{ }~~\&~
Werner Rodejohann$^b$\thanks{email: \tt werner$\_$rodejohann@ph.tum.de} 
\\\\
$^a${\normalsize \it Theoretical Physics, University of Oxford}\\
{\normalsize \it 1 Keble Road, Oxford OX1 3NP, 
UK}\\ \\
$^b${\normalsize \it Physik--Department, Technische Universit\"at M\"unchen,}\\
{\normalsize \it  James--Franck--Strasse, D--85748 Garching, Germany}
}
\date{}
\maketitle
\thispagestyle{empty}
\vspace{-0.8cm}
\begin{abstract}
\noindent 
We consider the flavor symmetry $L_\mu - L_\tau$ for the neutrino mass 
matrix. 
The most general neutrino mass matrix conserving $L_\mu - L_\tau$ 
predicts quasi--degenerate neutrino masses with one
maximal and two zero mixing angles. 
The presence of $L_\mu - L_\tau$ can also be motivated by the 
near--bimaximal form of the neutrino mixing matrix. 
Furthermore, it is a special case of $\mu \tau$ symmetric mass matrices. 
Breaking the flavor symmetry by adding a small flavor--blind term 
to the neutrino mass matrix and/or by applying 
radiative corrections is shown to reproduce the observed 
neutrino oscillation phenomenology. 
Both the normal and inverted mass ordering can be accommodated 
within this scheme. Moderate cancellation 
for neutrinoless double beta decay is expected. 
The observables $|U_{e3}|^2$ and $| 1/2 - \sin^2\theta_{23}|$ 
are proportional to the inverse of the fourth power of the common 
neutrino mass scale. 
We comment on whether the atmospheric neutrino mixing is 
expected to lie above or below $\pi/4$. We finally present a 
model based on the see--saw mechanism which generates a  
light neutrino mass matrix with an (approximate) $L_\mu - L_\tau$
flavor symmetry. This is a minimal model with just one standard 
Higgs doublet and three heavy right--handed neutrinos. It  
needs only small values for the soft $L_\mu - L_\tau$ breaking 
terms to reproduce the phenomenological viable mass textures analyzed.

\end{abstract}

\newpage

\section{\label{sec:intro}Introduction}
The structure of the neutrino mixing matrix is seen to be 
remarkably different from the quark mixing matrix. Within the standard 
parametrization of the PMNS \cite{PMNS} 
mixing matrix 
\bea \label{eq:Upara}
U = \left( \bad 
c_{12} c_{13} & s_{12} c_{13} & s_{13} 
\\[0.2cm] 
-s_{12} c_{23} - c_{12} s_{23} s_{13} 
& c_{12} c_{23} - s_{12} s_{23} s_{13}
& s_{23} c_{13} \\[0.2cm] 
s_{12} s_{23} - c_{12} c_{23} s_{13} 
& 
- c_{12} s_{23} - s_{12} c_{23} s_{13}
& c_{23} c_{13}\\ 
               \ea   \right) 
\, , 
\eea
where $c_{ij} = \cos\theta_{ij}$, $s_{ij} = \sin\theta_{ij}$ and 
the three physical phases were omitted, the following results emerged: 
\begin{itemize}
\item a small and possibly zero $\theta_{13} \simeq U_{e3}$ 
obtained from the results of the CHOOZ and Palo Verde experiments
\cite{ue3}; 
\item a very large and possibly maximal $\theta_{23}$ associated with a 
mass--squared difference $\dma \simeq 2 \cdot 10^{-3}$ eV$^2$ 
needed to explain the atmospheric neutrino data \cite{atm} and 
data from the K2K long baseline accelerator experiment \cite{k2k};
\item a large but non--maximal $\theta_{12}$  associated with a 
mass--squared scale $\dms \simeq 8 \cdot 10^{-5} $ eV$^2$ 
coming from the results on solar neutrinos \cite{sol} 
and the KamLAND reactor antineutrino experiment \cite{kl766}. 
\end{itemize}
This mixing structure has to be contrasted with the 
CKM matrix, whose structure is given 
in zeroth order by the unit matrix. 

In this letter we will consider the possibility that the unusual 
and unexpected structure of neutrino mixing is the 
consequence of a flavor symmetry acting on the neutrino mass matrix. 
In the basis in which the charged lepton mass matrix is real and diagonal, 
the neutrino mass matrix is defined as 
\be \label{eq:mnu}
m_\nu = U \, m_\nu^{\rm diag} \, U^T ~, 
\ee
where $m_\nu^{\rm diag}$ is a diagonal matrix containing the three 
neutrino masses $m_{1,2,3}$, for which the normal and inverted mass ordering 
are allowed. Those two possibilities correspond 
to $m_3 > m_2 > m_1$ and $m_2 > m_1 > m_3$, respectively. 
The fact that $\dma \gg \dms$, together with the limit on 
neutrino masses of order eV \cite{lisi}, implies that three extreme
kinds of mass spectra are possible:
\be \label{eq:mass}
\baz 
\mbox{ normal hierarchy:} 
& |m_3| \simeq \sqrt{\dma} \gg |m_{2}| \simeq \sqrt{\dms} \gg |m_1|~,\\[0.3cm]
\mbox{ inverted hierarchy:} 
& |m_2| \simeq |m_1| \simeq \sqrt{\dma} \gg |m_{3}| ~,\\[0.3cm]
\mbox{ quasi--degeneracy:} 
& m_0 \equiv |m_3| \simeq |m_2| \simeq |m_1|  \gg \sqrt{\dma} ~.
\ea 
\ee

\noindent 
The specific flavor symmetry we consider in this paper 
is $L_\mu - L_\tau$ which gives rise to a quasi--degenerate spectrum.
In general, a matrix strictly 
conserving $L_\mu - L_\tau$ will have the form 
\be \label{eq:mnu0}
m_\nu = m_0 \left( 
\bad 
c & 0 & 0 \\[0.3cm]
\cdot & 0 & s \\[0.3cm]
\cdot & \cdot & 0 
\ea 
\right)~,
\ee 
with a common scale $m_0$, $c = \cos \theta$ and $s = \sin \theta$. 
The matrix which diagonalizes $m_\nu $ is given by
\be
U = \left( 
\bad 
1 & 0 & 0 \\[0.3cm]
0 & \sqrt{\frac{1}{2}} & \sqrt{\frac{1}{2}} \\[0.3cm]
0 & -\sqrt{\frac{1}{2}} & \sqrt{\frac{1}{2}}
\ea 
\right)~,
\label{eq:mixmat0}
\ee 
independent of $\theta$. The mass eigenvalues are $c~m_0$ and $\pm s~m_0$. 
We see that for 
values of $\theta$ far from 0 or $\pi/2$, 
the three mass eigenstates 
are of the same order. 
Thus we obtain a quasi--degenerate neutrino mass scheme. 
This mass scheme is expected to be more easily testable than the 
hierarchical ones.  
Furthermore, we already have one maximal mixing angle, which can be 
identified with the atmospheric angle $\theta_{23}$ whose value is 
experimentally known to be close to maximal. Moreover, we also have 
one zero mixing which is compatible with 
the experimental upper limit on $\theta_{13}$.
The third mixing angle, which is the solar mixing 
angle $\theta_{12}$, comes out to be zero. 
This of course is in conflict with the current data from solar and 
KamLAND experiments. We shall see later that the non--zero 
value of $\theta_{12}$ could be
associated with order one numbers stemming from 
the breaking of the flavor symmetry under consideration.\\

We can also motivate the choice of $L_\mu - L_\tau$ flavor symmetry 
starting from an assumed bimaximal structure for the neutrino mixing.
The experimentally observed 
``bi--large'' structure of the mixing matrix can lead one to 
assume the bimaximal neutrino mixing scheme \cite{bimax} 
as a zeroth order approximation.  
Various proposals \cite{devbimax,PR,PR1} have been made in 
the literature to 
deviate the mixing from bimaximal in order to 
reproduce the observed phenomenology. 
Bimaximal mixing corresponds to 
$\theta_{12} = \theta_{23} = \pi/4$ and $\theta_{13} = 0$. 
Hence, 
\be
\label{eq:Ubimax}
U^{\rm bimax} = 
\left(
\bad  
\frac{1}{\sqrt{2}} &  \frac{1}{\sqrt{2}} &  0 \\[0.3cm]
-\frac{1}{2} &  \frac{1}{2} &  \frac{1}{\sqrt{2}} \\[0.3cm]
\frac{1}{2} &  -\frac{1}{2} &  \frac{1}{\sqrt{2}} \\[0.3cm]
\ea 
\right)~.
\ee
For bimaximal mixing the following mass matrix is implied: 
\bea \label{eq:mnubimax}
m_\nu = \left( \bad 
\D A & B & - B \\[0.2cm] 
\D \cdot & \D D + \frac{A}{2} &  \D D - \frac{A}{2} \\[0.3cm]
\D \cdot & \cdot & \D D + \frac{A}{2} 
\ea   \right)~, 
\eea
where 
\be
A = \frac{m_1 + m_2}{2}~,~
B = \frac{m_2 - m_1}{2 \, \sqrt{2}}~
,~D = \frac{m_3}{2}~. 
\ee
In case of leptonic $CP$ conservation, 
different relative signs for the mass states are possible. 
We can derive now three very interesting special cases of the matrix 
Eq.\ (\ref{eq:mnubimax}), obtained for the three extreme mass hierarchies 
mentioned above: 
\begin{itemize}
\item For normal hierarchy, i.e., $m_3 \gg m_{2,1}$ we have 
\bea \label{eq:NHex}
m_\nu \simeq 
\D \frac{\sqrt{\dma}}{2} \, \left( \bad 
\D 0 & 0 & 0 \\[0.2cm] 
\D \cdot & \D 1  &  \D 1 \\[0.3cm]
\D \cdot & \cdot & \D 1 
\ea   \right)
~.
\eea
This matrix conserves the flavor symmetry $L_e$. 
It displays the well--known ``leading $\mu\tau$ block'' \cite{leadmt} 
structure and produces (similar to the $L_\mu - L_\tau$ case) 
maximal mixing in the 23 sector and 
zero mixing in the 12 and 13 sector.  
The zero entries are filled with terms of order 
$\sqrt{R} \equiv \sqrt{\dms/\dma}$ once $m_2$ is no longer neglected with 
respect to $m_3$. 
Note that the determinant of the $\mu \tau$ block has then to 
be small in order to generate large 12 mixing.
\item For inverted hierarchy and $m_2 = - m_1$ we find 
\be \label{eq:IHex}
m_\nu = \sqrt{\frac{\dma}{2}} \,  
\ba 
\left( 
\bad 
0 & 1 & -1 \\[0.3cm]
\cdot &  0 &  0 \\[0.3cm]
\cdot & \cdot & 0 \ea \right)~.
\ea
\ee
This mass matrix conserves 
the lepton charge $L_e - L_\mu - L_\tau$ \cite{lelmlt,PR,PR1,grimus} 
and has been considered by many authors. 
Note that it generates exact bimaximal 
mixing\footnote{Strictly speaking, 
a matrix conserving $L_e - L_\mu - L_\tau$ does not have to have 
non--zero entries of equal magnitude. Consequently, 
atmospheric mixing is not {\it predicted} to be maximal and only 
$\theta_{13} = 0$ and $\theta_{12} = \pi/4$ are predicted by the 
most general mass matrix conserving $L_e - L_\mu - L_\tau$, 
see, e.g., \cite{PR1}.}.
\item For quasi--degenerate neutrinos,
on which we will focus in this letter, if we have
$m_1 = m_2 = -m_3$, we get
\be \label{eq:LmLt}
m_\nu = m_0 
\left( 
\bad 
1 & 0 & 0 \\[0.3cm]
\cdot & 0 & -1 \\[0.3cm]
\cdot & \cdot & 0 \ea \right)~.
\ee
This matrix is a special case of the matrix Eq.\ (\ref{eq:mnu0}) 
conserving the lepton charge $L_\mu - L_\tau$ and generates 
maximal mixing in the 23 sector and zero mixing in the 12 and 13 sectors. 
The matrix (\ref{eq:LmLt}) has been considered in very few 
papers in the literature ---  
for instance in \cite{PetRam,foot}, or more recently in 
the framework of an $A_4$ symmetry in \cite{A4}.  
In this respect, it has also been shown that 
the flavor symmetry $L_\mu - L_\tau$ is (in the Standard Model (SM)) 
anomaly free and may be gauged \cite{gauge}. 
To be precise, either $L_\tau - L_e$, $L_e - L_\mu$ or 
$L_\mu - L_\tau$ can be chosen to be gauged. As argued here, 
$L_\mu - L_\tau$ emerges as the phenomenologically most viable 
candidate among the three possibilities. 

\end{itemize}

\noindent 
In general there are 9 possible Abelian 
flavor symmetries for three active neutrinos: $L_e$, 
$L_\mu$, $L_\tau$ using only one flavor; $L_e - L_\mu$, $L_e - L_\tau$ 
and $L_\mu - L_\tau$ using two flavors; 
$L_e - L_\mu - L_\tau$, $L_e + L_\mu - L_\tau$ and $L_e - L_\mu + L_\tau$ 
using three flavors. We have shown that among these 9 candidates,
bimaximal neutrino mixing as a zeroth order 
approximation selects three of them:
$L_e$ (normal hierarchy), 
$L_e - L_\mu - L_\tau$ (inverted hierarchy) and 
$L_\mu - L_\tau$ (quasi--degeneracy).

The $L_e - L_\mu - L_\tau$ ($L_e$) flavor symmetry is 
often argued as the Abelian symmetry of the underlying theory 
which produces inverted (normal) hierarchy. 
Along the same lines, we champion here the case of  
$L_\mu - L_\tau$ as the underlying symmetry of the theory 
which generates 
a quasi--degenerate spectrum for the neutrinos.\\

There is yet another motivation for $L_\mu - L_\tau$: recently, the 
presence of a $\mu\tau$ symmetry in the neutrino mass matrix has been put 
forward to explain maximal $\theta_{23}$ and zero $\theta_{13}$ \cite{mutau}.
Such a mass matrix reads 
\be \label{eq:mutau}
m_\nu = 
\left( 
\bad
A & B & B \\[0.3cm]
\cdot & D & E  \\[0.3cm]
\cdot & \cdot & D 
\ea
\right)~ 
\ee
and obviously predicts  $\theta_{23} = \pi/4$ and zero $\theta_{13}=0$. 
The special case $A = D = E = 0$ corresponds to a special 
case of $L_e - L_\mu - L_\tau$ (namely giving exact bimaximal mixing), 
whereas $D = E = 0$ yields a matrix conserving $L_e$. 
Finally, setting $B = D = 0$ results in a mass matrix of the form 
of Eq.\ (\ref{eq:mnu0}), i.e., conserving $L_\mu - L_\tau$.\\

We finally note that a discrete phase transformation 
$e \rightarrow e$, $\mu \rightarrow i \mu$ and $\tau \rightarrow -i \tau$ 
will generate a neutrino mass matrix with the same structure as 
Eq.\ (\ref{eq:mnu0}). In this paper, however, we want to focus on simple 
Abelian $U(1)$ flavor symmetries.

\section{\label{sec:Lmt}The flavor symmetry $L_\mu - L_\tau$:
General Considerations} 

As mentioned above, the most general matrix  
preserving $L_\mu - L_\tau$ (cf.\ Eq.\ (\ref{eq:mnu0})) is diagonalized by
the mixing matrix given by Eq.\ (\ref{eq:mixmat0}). 
The mass eigenvalues  $c~m_0$ and $\pm s~m_0$ are of the same order 
when the two non--vanishing entries in the mass matrix,  
$m_{ee} = m_0 \, \cos \theta$ and $m_{\mu \tau} = m_0 \, \sin \theta$, 
are of the same order. 
There is only one non--zero 
$\Delta m^2 = m_0^2 \, \cos 2\theta$, which would 
disappear when $\theta = \pi/4$, i.e., for an additional 
symmetry that would lead to treating the $ee$ and 
$\mu \tau$ elements equally. 
This non--vanishing $\Delta m^2$ corresponds to the 
{\it solar} mass--squared difference. The atmospheric 
mass--squared difference which is predicted to be zero is 
expected to be generated by breaking the $L_\mu - L_\tau$ symmetry. 
Therefore any perturbation of the 
$L_\mu - L_\tau$ symmetry has to make the a priori zero $\dma$ larger than 
the a priori non--vanishing $\dms = m_0^2 \, \cos 2\theta$. Hence we can 
expect that to explain the neutrino oscillation data, 
$\cos 2\theta$ should be close to zero, because then the 
``flipping'' of the magnitudes of the solar and atmospheric 
$\Delta m^2$ will be easier. 
Moreover, the arguments based on the near--bimaximality of the mixing 
structure given above also indicate that the two entries in $m_\nu$ 
should be very similar (cf.\ Eq.\ (\ref{eq:LmLt})).

Regarding the common mass scale $m_0$, 
cosmological limits on neutrino masses bound the quantity 
$\Sigma \simeq  3 m_0$ to be less than 0.4 to 2 eV, depending on the 
priors and the data set used to obtain the limit (see e.g.\ 
\cite{lisi,steen,nunew} for a discussion). 
Typically, $m_0 \gs 0.2$ eV --- being on the edge of 
being ruled out by the strictest bounds from observations ---
is a value required to lead to a quasi--degenerate scheme for neutrino 
masses. Leaving out the 
Ly--$\alpha$ data, whose systematics seem not to be fully understood, 
limits of $\Sigma \ls 1.5$ eV are obtained, which leaves still enough 
room for the possibility of quasi--degenerate neutrinos. 
Furthermore, the limit 
on the effective mass 
measurable in \onbb \cite{0vbb} experiments, which is the 
$ee$ entry of the neutrino mass matrix, is given by 
$\meff < 0.89$ eV at $1\sigma$ \cite{lisi}, 
taking into account the uncertainty in the nuclear 
matrix element calculations. Due to possible cancellations in this 
element \cite{canc}, the common mass scale might be larger 
by a factor $\simeq 1/(1 - \tan^2 \theta_{\odot}) \ls 2$.

To estimate in a bottom--up approach 
the realistic form of $m_\nu$ corresponding to an approximate 
$L_\mu - L_\tau$ symmetry, it is very convenient to 
parameterize the deviations from bimaximal mixing with a small 
parameter $\lambda$, defined via \cite{ichPRD}
\be
U_{e2} = \sqrt{\frac{1}{2}} (1 - \lambda )~.
\ee 
Typical best--fit points correspond to $\lambda \simeq 0.22$ \cite{ichPRD}, 
which is remarkably close to the Cabibbo angle\footnote{Note that this 
implies the most interesting equation $\theta_C + \theta_\odot = \pi/4$, 
which can be realized, e.g., in the framework of quark--lepton symmetry, 
e.g., if $U_\nu$ is bimaximal, $U_{\rm up} = \mathbbm{1}$, 
$U_{\rm down} = U_{\rm CKM}$ and $U_{\rm lep} = U_{\rm down}$ \cite{QLC}.} 
$\theta_C$.  
One can further describe the expected deviations from zero $|U_{e3}|$ and 
$\theta_{23} = \pi/4$ by writing $|U_{e3}| = A \, \lambda^n$ as well as 
$U_{\mu 3} = \sqrt{1/2} (1 - B \, \lambda^m)$ with integer $n,m$ and 
$A,B$ of order 1 \cite{ichPRD}. Regarding the masses, we can write 
(for the normal ordering) $m_3 = m_0$, $m_2 = m_0\sqrt{1 - \eta}$ 
and $m_1 = m_0\sqrt{1 - \eta~(1 + C~\lambda^2)}$, where we used that 
$R = C~\lambda^2$ \cite{ichPRD}, with $C$ of order one. 
The small parameter $\eta$ 
is defined as $\eta \equiv \dma/m_0^2$, for which typically 
$\eta \ls \lambda^2$ holds when $m_0 \gs 0.2$ eV. In case of lower values 
of $m_0 \ls 0.1$ eV, we can have $\eta \gs \lambda$.  
For both $|U_{e3}|$ and $\theta_{23}$ 
being very close to their current $3\sigma$ bounds, which correspond 
to $|U_{e3}|^2 \sim 0.05$ and $\sin^2 2 \theta_{23} \sim 0.9$ we have 
$m = n = 1$ and the mass matrix (\ref{eq:LmLt}) is modified to 
\bea \label{eq:mnu11} \small 
\frac{\D m_\nu}{\D m_0} \simeq   
\left(  \bad 
1 - 2 \, A^2 \,  \lambda^2 - \frac{\D \eta}{\D 2} & 
- \sqrt{2} \, A \, \lambda + \sqrt{2} \, A \, B \, \lambda^2 & 
- \sqrt{2} \, A \, \lambda - \sqrt{2} \, A \, B \, \lambda^2 \\[0.3cm] 
\cdot &  2 \, B \, \lambda - B^2 \, \lambda^2 - \frac{\D \eta}{\D 4} & 
-1  + (A^2 + 2 \, B^2 )  \, \lambda^2 + \frac{\D \eta}{\D 4} \\[0.3cm]
\cdot & \cdot &  - 2 \, B \, \lambda + (2 \, A^2 + B^2) 
\, \lambda^2 - \frac{\D \eta}{\D 4}
\ea 
\right) 
\eea 
plus terms of order ${\cal O}(\eta\,\lambda , \lambda^3)$ and higher. 
The parameter $C$ classifying the difference between $R$ and the deviation 
from maximal solar neutrino mixing does not appear at the order given above. 
As seen from Eq.\ (\ref{eq:mnu11}), corrections required to 
reproduce the observed phenomenology are 
sizable for the zero entries and small for the entries equal to 1. 
Different corrections, e.g.\ for $m = 2$ and $n = 3$ 
can be obtained by replacing 
$A$ with $A~\lambda^2$ and $B$ with $B~\lambda$. Taking the inverted ordering 
for the neutrino masses will make $\eta$ disappear in the $ee$ entry 
and have the term proportional to $\eta$ change its sign in the $\mu\mu$ 
and $\tau\tau$ entries. For values corresponding to smaller 
$|U_{e3}|^2 \ls 0.01 $ and larger $\sin^2 2 \theta_{23} \gs 0.95$ 
all corrections to the zeroth order mass matrix Eq.\ (\ref{eq:LmLt}) will 
be quadratic.

\section{\label{sec:breaki}Generating successful phenomenology
from $L_\mu - L_\tau$}

We have seen above that exact $U(1)$ flavor symmetry $L_\mu - L_\tau$ predicts
exact maximal mixing for atmospheric neutrinos and zero mixing 
for the CHOOZ mixing angle, which is completely consistent with data.
However, the solar mixing angle as well as 
the neutrino mass splittings are inconsistent with the solar 
and atmospheric data. To generate correct values for these observables, 
we have to break the $L_\mu - L_\tau$. 
In the first of the next two Subsections we break this symmetry 
by adding a small ``democratic'' perturbation 
to the zeroth order mass matrix. Simple formulae which are able to express 
interesting correlations between the observables are possible to 
write down in this case. 
In the second Subsection we add a 
random perturbation matrix in which each of the elements could take any 
(small) value. We show, using approximate analytical 
expressions as well as exact numerical results, that the global 
oscillation data are completely consistent with approximate 
$L_\mu - L_\tau$ flavor symmetry in the neutrino sector. 
Though the presence of random perturbation to Eq.\ (\ref{eq:mnu0}) 
seems more likely than a purely democratic flavor--blind correction, the 
results from the two approaches turn out not to differ drastically, 
which is a reassuring fact.  

\subsection{\label{sec:dem}Democratic perturbation plus radiative corrections}
A very simple perturbation of the zeroth order mass matrix 
Eq.\ (\ref{eq:mnu0}) is obtained by adding a small and purely democratic 
correction to the mass matrix. The new mass matrix reads 
\be \label{eq:mnudem}
m_\nu = m_0 \left[ \left( 
\bad 
c & 0 & 0 \\[0.3cm]
\cdot & 0 & s \\[0.3cm]
\cdot & \cdot & 0 
\ea 
\right) 
+ \epsilon 
\left( 
\bad 
1 & 1 & 1 \\[0.3cm]
\cdot & 1 & 1 \\[0.3cm]
\cdot & \cdot & 1 
\ea 
\right) \right] ~.
\ee 
While it turns out that with this matrix alone no successful 
phenomenology can be generated (because of the implied 
vanishing solar neutrino mixing), 
radiative corrections \cite{RGE} 
from the scale at which this matrix is generated down to low energy 
are seen to do the job. 
Effects of radiative corrections can be estimated by multiplying 
the $\alpha \beta$ element of $m_\nu$ with a term 
$(1 + \delta_\alpha) \, (1 + \delta_\beta)$, where 
\be \label{eq:RGE}
\delta_\alpha = c \, \frac{m_\alpha^2}{16 \, \pi^2 \, v^2} 
\, \ln \frac{M_X}{m_Z}~.
\ee   
Here $m_\alpha$ is the mass of the corresponding charged lepton, 
$M_X \simeq 10^{16}$ GeV and $c = -(1 + \tan^2 \beta)$ (3/2) in case of the 
MSSM (SM). We will see that
in the SM, the induced corrections are found to be insufficient.
In fact we will see that we 
need large $\tan\beta$ to explain the experimental data. 

\begin{table}[t]
\begin{center}
\begin{tabular}{|c|c|c|} 
\hline
$L'$ & matrix & extra requirement \\ \hline \hline
$\ba L_e \\ \mbox{\small normal hierarchy} \ea$ & 
$\left( \bad 0 & 0 & 0 \\ \cdot & a & b \\ \cdot & \cdot & d \ea \right) $
& $\ba a \simeq d~(\leftrightarrow \mbox{maximal } \theta_{23}) 
\\ \mbox{small } ad - b^2~(\leftrightarrow \mbox{large } \theta_{12})\ea$  
\\ \hline 
$\ba L_e - L_\mu - L_\tau \\ \mbox{\small inverted hierarchy} \ea $ & 
$\left( \bad 0 & a & b \\ \cdot & 0 & 0 \\ \cdot & \cdot & 0 \ea \right) $
& $ \ba a \simeq b~(\leftrightarrow \mbox{maximal } \theta_{23}) 
\\ \mbox{needs } U_{\rm lep} \mbox{ or strong breaking}\ea$ \\ \hline 
$\ba L_\mu - L_\tau \\ \mbox{\small quasi--degeneracy} \ea $ & 
$\left( \bad a & 0 & 0 \\ \cdot & 0 & b \\ \cdot & \cdot & 0 \ea \right) $ 
& $ a \simeq -b~(\leftrightarrow R < 1) $ \\ \hline
\end{tabular}
\caption{\label{tab:tab1}The three Abelian $U(1)$ flavor symmetries implied 
by data, the implied neutrino mass matrix and the extra requirements 
(apart from soft breaking) in 
order to achieve successful phenomenology.}
\end{center}
\end{table}
Sticking to positive $\epsilon$, numerically it turns out 
that $\theta \simeq 3\pi/4$ is required to generate neutrino masses and 
mixings in accordance with the recent data. Then the normal mass ordering is 
predicted. For $\theta \simeq 7\pi/4$ (or alternatively negative $\epsilon$) 
we end up with the inverted ordering. This confirms our earlier suspicion 
that $\cos 2 \theta$ 
should be close to zero to allow for a ``flip'' in the 
magnitude of the original \dms{} and $\dma$ predicted by the 
mass matrix given in Eq.\ (\ref{eq:mnu0}). This implies that 
we need in addition another symmetry which will naturally explain 
why the $ee$ element of the zeroth order matrix (\ref{eq:mnu0})
should be equal to the $\mu\tau$ element. However we have encountered 
situations like this before. For instance, for 
the case of $L_e - L_\mu - L_\tau$ symmetry, maximal mixing 
for the atmospheric is not the most general prediction of the 
theory and one requires the two 
non--zero entries in Eq.\ (\ref{eq:IHex}) to be 
equal (see \cite{PR1} and references therein).
In case of $L_e$ conservation, 
all three non--zero entries (the $\mu\tau$ block) are 
required to have roughly equal magnitude in order 
to accommodate maximal atmospheric neutrino mixing and, 
after breaking (via a small 23 sub--determinant) 
large solar neutrino mixing. 
In the case of $L_\mu - L_\tau$ it is the small observed value for the 
ratio $R$ of the $\Delta m^2$ that 
forces the $ee$ and $\mu \tau$ entries to
have nearly equal magnitude. 
It seems therefore that a simple flavor symmetry has to be 
spiced up with 
additional symmetries to explain the global experimental data\footnote{For 
$L_e - L_\mu - L_\tau$ one requires in addition large contributions  
from charged lepton mixing in order to generate a viable
neutrino phenomenology \cite{PR1}. Alternatively, the symmetry breaking 
terms in the mass matrix should be of the same order as the terms allowed 
by the symmetry \cite{grimus}.}. Table \ref{tab:tab1} summarizes the 
situation.

For the sake of obtaining approximate analytic expressions for the 
mass and mixing parameters we
neglect $\delta_{e,\mu}$ and diagonalize the matrix 
\be
m_\nu \simeq m_0  \left( 
\bad 
c + \epsilon &  \epsilon & \epsilon~(1 + \delta_\tau) \\[0.3cm]
\cdot & \epsilon & (s + \epsilon)~(1 + \delta_\tau)\\[0.3cm]
\cdot & \cdot & \epsilon~(1 + 2~\delta_\tau) 
\ea 
\right) ~.
\ee
For $\theta \simeq (2n + 1)\pi/4$ (with integer $n$)  
in leading order for the three mass states 
$|m_3| \simeq |m_2| \simeq |m_1|$. 
For $\theta = 3\pi/4$ we get normal ordering of the mass spectrum with
\be 
m_3 \simeq \frac{m_0}{\sqrt{2}}(1
+ \delta_\tau + 2~\sqrt{2}~\epsilon)~~,~~
m_2 \simeq -\frac{m_0}{\sqrt{2}}(1 + \delta_\tau)~~,~~
m_1 \simeq -\frac{m_0}{\sqrt{2}}~\left(1 - \sqrt{2}~\epsilon\right)~,
\ee
which leads to 
\be \label{eq:massNH} 
\dma \simeq m_0^2~2\sqrt{2}~\epsilon~~\mbox{ and }~~ 
\dms \simeq m_0^2~(\sqrt{2}~\epsilon + \delta_\tau)~.
\ee
From these numerically not very precise 
expressions one can nevertheless see that the value of 
$\dma$ is determined by $\epsilon$ while the $\dms$ 
depends on both $\delta_\tau$ and 
$\epsilon$. We note that in order to produce $\dms \ll \dma$
we need opposite signs for $\delta_\tau$ and $\epsilon$. 
In fact, the relation $\epsilon \sim -\delta_\tau/\sqrt{2}$ is needed to 
generate a small ratio $R$ of the solar and atmospheric $\Delta m^2$. 
From $\dma \simeq 2 \cdot 10^{-3}$ eV$^2$ and a common neutrino mass scale 
of $m_0 = 0.1~(0.5)$ eV, we can estimate $\epsilon \simeq 0.07~(0.003)$. 
Since $\delta_\tau $ should be of the same order, we demand that
$\tan \beta$ has to be larger than 10, as can be seen from 
Eq.\ (\ref{eq:RGE}). For the lowest possible mass of $m_0 \simeq 0.05$ eV 
we find $\epsilon \simeq 0.28$. Sticking to reasonable values of 
$\tan^2 \beta \ls 10^3$ and 
noting that $\epsilon \sim |\delta_\tau|$, we can estimate however that 
 $\epsilon \simeq 0.1$ is a more reasonable upper value. 

The effective mass for the $0\nu\beta\beta$ process is given by 
\be \label{eq:NHmeff}
\meff = m_0 \left| \cos \theta + \epsilon \right| \simeq 
m_0 \left( \sqrt{\frac{1}{2}} - \epsilon \right)~,  
\ee
so that only modest cancellation (i.e., maximal 15$\%$ for the 
largest $\epsilon \simeq 0.1$) is predicted.  
The solar neutrino mixing angle is given by 
\be
\tan^2 \theta_{12} \simeq 2~\left|\frac{\epsilon}{\delta_\tau}\right|~.
\ee
Recalling that the zeroth order mass matrix Eq.\ (\ref{eq:mnu0}) predicts 
zero $\theta_{12}$, we see that large solar mixing is indeed 
associated with a ratio of two numbers of almost equal magnitude
which are responsible for the breaking of the symmetry. 

For the currently unknown mixing parameters $U_{e3}$ and 
$\sin^2  \theta_{23}$ we have 
\be \label{eq:NH1}
U_{e3} \simeq \epsilon~(1 - \epsilon/\sqrt{2})~~
\mbox{ and } 
\sin^2  \theta_{23} \simeq \frac{1}{2} - 
\frac{\delta_\tau \epsilon}{\sqrt{2}} > \frac{1}{2}~. 
\ee
We see that $U_{e3}$ is directly proportional to 
the symmetry
breaking parameter $\epsilon$ and can estimate an allowed range of 
\be \label{eq:ue3NH}
U_{e3} \simeq \frac{\dma}{m_0^2~2\sqrt{2}} \simeq 
0.003 \ldots 0.1 ~,
\ee
where we varied $m_0$ between 0.05 and 0.5 eV. 
With the form of $\dma$ given in Eq.\ (\ref{eq:massNH}) plugged in 
Eq.\ (\ref{eq:ue3NH}), the value of $U_{e3}$ is seen to be 
inversely proportional to the square of the common mass scale. 
With $\delta_\tau \sim -\epsilon$ we furthermore see that the deviation 
from $\sin^2  \theta_{23} = 1/2$ is proportional to $U_{e3}^2$ 
and therefore inverse proportional to the fourth power of the 
common mass scale. 
We can expect 
\be \label{eq:saNH}
\sin^2  \theta_{23} - \frac{1}{2}  \simeq 
\frac{\sqrt{2}}{\tan^2 \theta_{12}}~U_{e3}^2 \simeq 
\left(\frac{\dma}{m_0^2~2\sqrt{2}}\right)^2 \ls 0.01~.
\ee
Such small deviations from maximal mixing are very hard, 
if not impossible, 
to measure \cite{atmnew} experimentally. 
As can be seen, atmospheric mixing lies 
on the ``dark side'' of the parameter space. 
There are physical observables sensitive to the difference 
$\theta_{23} > (<) \, \pi/4$ \cite{atmnew,lida}, though the 
deviations of $\theta_{23}$ from maximality into the dark side that 
we obtain in this Subsection are too small
to be observed experimentally.

Given that cosmological 
observations will restrict or measure $\Sigma \simeq 3 m_0$, 
we can deduce from Eqs.\ (\ref{eq:NHmeff},\ref{eq:NH1}) that 
\be
\Sigma \simeq 3\sqrt{2}~ \meff~(1 - \sqrt{2}~U_{e3})~,
\ee
linking cosmology with \onbb$\!\!$.\\

In case of the inverted mass ordering for 
the neutrino, which is obtained 
for $\theta \simeq 7 \pi/4$, we find very similar expressions. 
However, numerically it turns out that values of $\theta$ slightly lower than 
$7 \pi/4$ are preferred, making a precise analytical calculation of the  
observables very difficult. 
Nevertheless, the form of $U_{e3}$ and $\sin^2  \theta_{23}$ 
can be estimated rather reliably for $\theta = 7 \pi/4$: 
\be \label{eq:IH1}
U_{e3} \simeq \epsilon~(1 + \epsilon/\sqrt{2})~~
\mbox{ and } 
\sin^2  \theta_{23} \simeq \frac{1}{2} - \frac{\epsilon^2}{2} + 
\frac{\delta_\tau \epsilon}{\sqrt{2}} < \frac{1}{2}~,
\ee
that is, we expect marginally larger values for
$U_{e3}$ compared to the case for normal mass ordering 
(cf.\ Eq.\ (\ref{eq:NH1})). 
We also find that atmospheric neutrino mixing lies 
on the ``light side'' of the parameter space, though it is 
probably still too small to be observable. 
The effective mass for $0\nu\beta\beta$ decay for 
the case of 
inverted ordering is given by $\meff \simeq m_0 (\sqrt{1/2} + \epsilon)$, 
slightly larger than for the normal ordering. 
Finally, the value  of $\theta$ slightly smaller than $7 \pi/4$ implies that 
$|\delta_\tau|$ and therefore also $\tan^2 \beta$ are slightly smaller 
than in case of the normal mass ordering. Due to this fact, the deviation 
from $\pi/4$ of $\theta_{23}$ is smaller than in case of the normal 
mass ordering discussed above. 

\begin{figure}[t]\vspace{-2cm}
\begin{center}\vspace{-2cm}
\epsfig{file=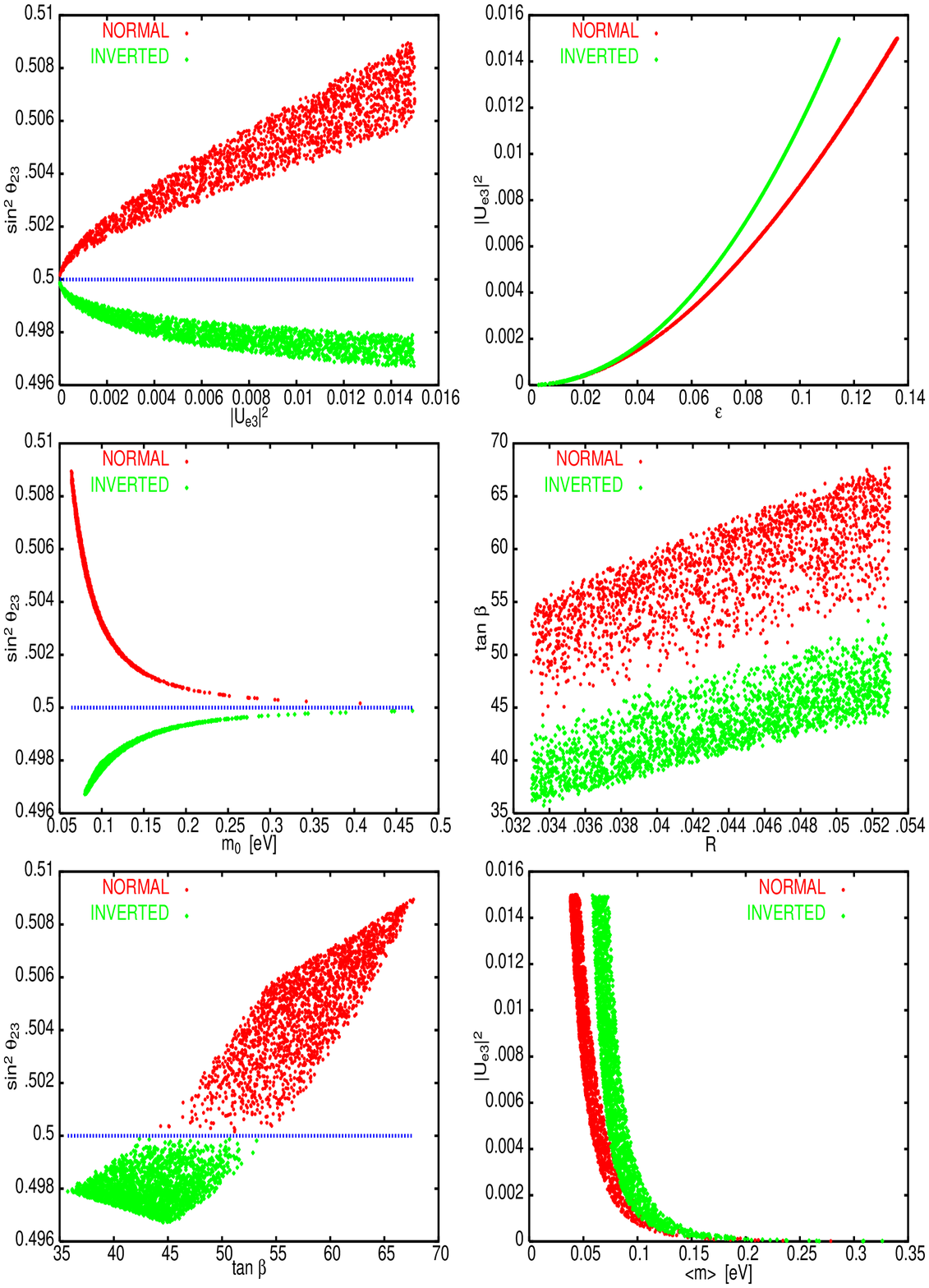,width=19cm,height=24cm}
\vspace{-2cm}
\caption{\label{fig:fig1}Scatter plots of some of the correlations 
of the parameters and observables resulting from Eq.\ (\ref{eq:mnudem}) plus 
applying radiative corrections.}
\end{center}
\end{figure}

We plot in Fig.\ \ref{fig:fig1} some of the resulting correlations 
of the parameters and observables obtained from an exact numerical
analysis, for both the normal and inverted 
hierarchy. The density of points contains no information, it is rather the 
envelope of the points which defines the physics. 
To produce the plots, we demanded 
the following values for the observables \cite{atm,sol,kl766,kl766us}: 
\begin{eqnarray} \label{eq:data}
\tan^2 \theta_{12} & = & 0.34 \ldots 0.44 \; ,
\nonumber \\
|U_{e3}|^2 & \le & 0.015 \; ,
\nonumber \\
\sin^2 2 \theta_{23} & \ge & 0.95 \; ,
\nonumber \\
R_\nu & \equiv & \frac{\Delta m^2_{\odot}}{\Delta m^2_{\rm A}}
= 0.033 \ldots 0.053 \; .
\end{eqnarray}
Furthermore, since only the ratio of the $\Delta m^2$ is required to be 
correctly reproduced, we restricted 
the common neutrino mass to be below 0.5 
eV. The small parameter $\epsilon$ is bounded from above by $1/\sqrt{10}$.
It can be seen from the Figure that, 
\begin{itemize}
\item the dependence of $U_{e3}$ and $\sin^2 \theta_{23}$ as given 
in Eqs.\ (\ref{eq:NH1},\ref{eq:IH1}) is correct. Zero $U_{e3}$ implies 
$\theta_{23} = \pi/4$ and vice versa;  
\item $U_{e3}$ and $\sin^2 \theta_{23}$ follow the functional behavior 
according to Eqs.\ (\ref{eq:ue3NH},\ref{eq:saNH}) to a good precision. Hence 
$U_{e3} \propto 1/m_0^2$ and $|1/2 - \sin^2 \theta_{23}| \propto 1/m_0^4$; 
\item atmospheric neutrino mixing lies on the ``dark side'' (``light side'') 
for the normal (inverted) mass ordering; 
\item the deviation from maximal $\theta_{23}$ can be up to two times 
larger in case of the normal ordering; 
\item neutrinoless double beta decay is driven by an effective 
mass between 0.05 and 0.35 eV (for a given $U_{e3}$ somewhat larger 
values are expected in case of the inverted ordering) and should 
therefore be observable (at the latest) in  next generation 
experiments \cite{0vbb};  
\item the common neutrino mass scale $m_0$ is between 0.05 and 0.5 eV and 
can be probed by next generation cosmological observations \cite{steen} and 
perhaps by the KATRIN laboratory experiment \cite{KATRIN}; 
\item for the normal ordering we need $\tan \beta \simeq 35 \ldots 50$ 
and for the inverted ordering $\tan \beta \simeq 45 \ldots 65$
is required. 
\end{itemize} 
Especially the last issue of the relatively large values of $\tan \beta$ 
is interesting because many lepton flavor violating processes such as, e.g., 
$\mu \ra e \gamma$ have a strong dependence on this quantity and can be 
expected to be sizable. Moreover, an interesting difference with respect to 
the studies in \cite{A4} can be seen. In those works, the matrix 
Eq.\ (\ref{eq:mnu0}) with $c = s$ has been derived from $A_4$ symmetry 
and the {\it most general} 
radiative corrections, including slepton threshold effects, were applied to 
reproduce the correct neutrino phenomenology. 
As a result, values of $\tan \beta$ smaller than 8 were required in order 
to reproduce the observed neutrino phenomenology \cite{A4}. Hence, the 
large difference between the requisite values of $\tan \beta$ may serve as 
a tool to distinguish the approach in \cite{A4} and the one presented here.

We remark that the much discussed \cite{disc} claim of a 
possible evidence of neutrinoless double beta decay, 
corresponding to $\meff \simeq 0.1 \ldots 0.9$ eV \cite{klapdor}, 
can in principle be accommodated with the flavor symmetry and 
neutrino mass matrix under study. 

We can compare the neutrino oscillation observables obtained 
in this section with the Ansatz motivated by the near--bimaximal 
neutrino mixing which led to Eq.\ (\ref{eq:mnu11}). 
For values of $\epsilon \simeq 0.1$ we get
$m_0 \simeq 0.08$ eV from 
Eq.\ (\ref{eq:massNH}) and 
$U_{e3} \simeq 0.1$ from Eq.\ (\ref{eq:NH1}). 
Therefore in the parameterization 
introduced in \cite{ichPRD} and 
given by Eq.\ (\ref{eq:mnu11}), we have $n=1$. 
The near--maximality of atmospheric neutrino 
mixing means that $m=3$ and we can compare our results with 
Eq.\ (\ref{eq:mnu11}), when we replace $B$ with $B \lambda^2$ in that 
Equation (note also the relative factor $\sqrt{2}$ for the definition 
of $m_0$ in Eqs.\ (\ref{eq:mnu11}) and (\ref{eq:mnudem})). 
Hence, we can estimate $\epsilon \simeq \eta/4\sqrt{2}$ and 
$\lambda \simeq \epsilon/A$. The result $\epsilon \simeq \eta/4\sqrt{2}$ 
is also found for the case of very small $\epsilon \ls 0.01$, 
resulting in small $U_{e3}$ and rather large $m_0 \gs 0.2$ eV. As a 
consequence, $\eta \simeq \lambda^2 \sim \sqrt{8}~\epsilon$. Similar 
arguments apply in the next Subsection, when we allow 
random order one coefficients 
for the perturbation to Eq.\ (\ref{eq:mnu0}). 

A few words on the origin of the democratic perturbation is in order. 
If $\epsilon$ would come from Planck 
scale effects \cite{Planck}, then its 
size would naively be given by 
$m_0~\epsilon \simeq v^2/M_{\rm Pl} 
\simeq 2.5 \cdot 10^{-6}~(m_0/{\rm eV})$ eV, 
with the Planck mass $M_{\rm Pl} \simeq 1.2 \cdot 10^{19}$ GeV.  
Hence, we cannot generate the required values for the 
neutrino mixing observables. Low scale gravity would 
effectively replace $M_{\rm Pl}$ with some scale $M_{\tilde{X}}$ in the above 
considerations. If, say, 
$M_{\tilde{X}} = 10^{16}$ GeV and $m_0 = 0.5$ eV then we could obtain 
\dma{} in the right ballpark. Of course, again values of 
$\tan\beta \sim 10$ are required to generate a correct \dms.

\subsection{\label{sec:UK}Anarchical perturbation}

In the last Subsection we considered a purely democratic perturbation 
(see Eq.\ (\ref{eq:mnudem})) to the 
zeroth order mass matrix conserving $L_\mu - L_\tau$. 
In this section we relax this constraint 
and allow for random order one coefficients for 
every entry in the perturbation matrix: 
\be \label{eq:mnudem1}
m_\nu = m_0 \left[ \left( 
\bad 
c & 0 & 0 \\[0.3cm]
\cdot & 0 & s \\[0.3cm]
\cdot & \cdot & 0 
\ea 
\right) 
+ \epsilon 
\left( 
\bad 
a &  b & d \\[0.3cm]
\cdot & e & f \\[0.3cm]
\cdot & \cdot & g  
\ea 
\right) \right] ~,
\ee 
where $a,b,d,e,f,g$ are real, positive and of order one. 
Approximate solutions for the oscillation parameters in case 
of $\theta = 3 \pi/4$ (i.e., in case of the normal mass ordering) read: 
\be \label{eq:NH1a}
U_{e3} \simeq \frac{1}{2}~(b + d)~\epsilon~(1-\epsilon/\sqrt{2})~~
\mbox{ and } 
\sin^2  \theta_{23} \simeq \frac{1}{2} + \epsilon ~
\frac{e - g}{\sqrt{8}} ~.
\ee
There is now a term proportional to $\epsilon$ for $\sin^2  \theta_{23}$, 
which vanishes for the case of ``democracy'', $e = g$. 
Also, the previous expressions in 
Eq.\ (\ref{eq:NH1}) are reproduced when $b = d$. The expressions for the 
$\Delta m^2$ are complicated and depend on $\epsilon$ and all six 
coefficients. 
In case of the inverted mass ordering, which is again obtained for 
 $\theta \simeq 7 \pi/4$, we can estimate 
\be \label{eq:IH1a}
U_{e3} \simeq \frac{1}{2}~(b + d)~\epsilon~(1+\epsilon/\sqrt{2})~~
\mbox{ and }~~ 
\sin^2  \theta_{23} \simeq \frac{1}{2} - \epsilon ~
\frac{e - g}{\sqrt{8}} ~.
\ee
In contrast to the case treated in Section \ref{sec:dem}, it is no more 
possible to predict whether atmospheric mixing lies above or below 
$\pi/4$ because this depends crucially on the relative magnitude of the order 
one coefficients.
Comparing however the expressions for $\sin^2  \theta_{23}$ in case of normal 
and inverted mass ordering, we see that for $e > g$ and the 
normal (inverted) ordering, 
atmospheric neutrino mixing will be on the ``dark side'' (``light side''), 
i.e., $\theta_{23} > (<) \, \pi/4$. For $g > e$ the situation is vice versa. 
Note further that there are different order one 
coefficients for $U_{e3}$ and $\sin^2  \theta_{23}$, making it 
impossible to write down interesting correlations between the mixing 
observables. 
Due to the order one coefficients, we can however expect now broader 
allowed ranges of $U_{e3}$ and $|\sin^2  \theta_{23} - 1/2|$ than before.

\begin{figure}[t]\vspace{-2cm}
\begin{center}\vspace{-2cm}
\hspace{-2cm}
\epsfig{file=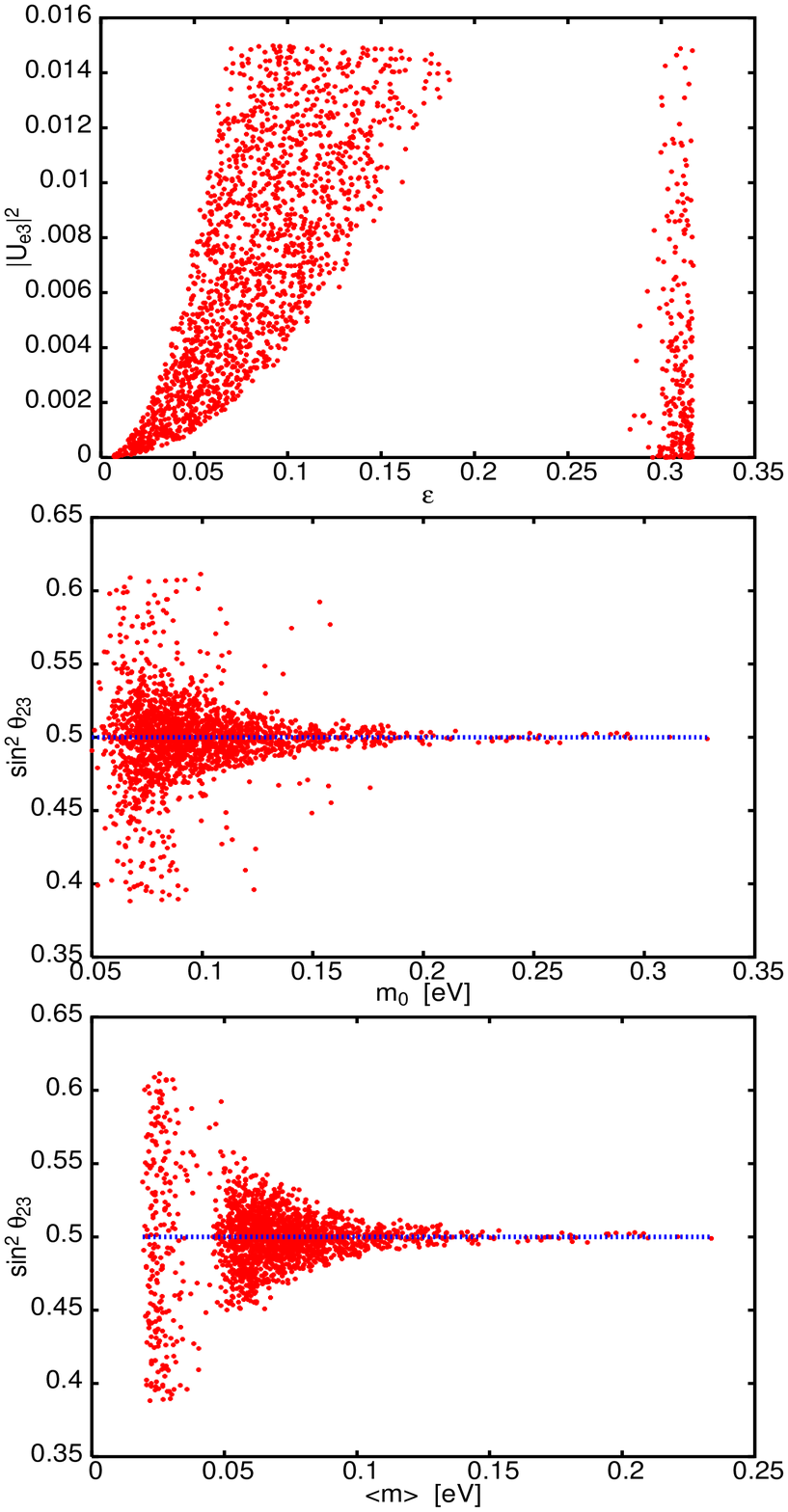,width=19cm,height=24cm}
\vspace{-1cm}
\caption{\label{fig:fig2}Scatter plots of some of the correlations 
of the parameters and observables resulting from Eq.\ (\ref{eq:mnudem1}). 
The plots are for the inverted mass ordering, the corresponding figure for the 
normal ordering löooks basically identical.}
\end{center}
\end{figure}

We plot in Fig.\ \ref{fig:fig2} some scatter plots of the 
observables, which are as before required to lie inside 
the ranges given in Eq.\ (\ref{eq:data}). 
All coefficients $a,b,d,e,f,g$ are varied between $1/\sqrt{3}$ 
and $\sqrt{3}$, $\epsilon$ is again bounded from above by $1/\sqrt{10}$.  
We find no significant difference between the normal and inverted 
mass ordering. As excepted, the ranges of the observables are broader than in 
the case of a purely democratic perturbation. There is also no upper limit 
on $\epsilon$ any more. Atmospheric neutrino mixing is described by either 
$\theta_{23} > \pi/4$ or $\theta_{23} < \pi/4$, where the deviation from 
maximal can be up to 20$\%$. It is also seen that --- as in the case 
of purely democratic perturbation --- the larger the neutrino mass 
scale, the closer $\theta_{23}$ is to $\pi/4$.

\section{\label{sec:model}A Simple Model}

In this Section we present
a very simple model based on the see--saw \cite{seesaw} 
mechanism which leads to a low energy mass matrix 
(approximately) conserving $L' \equiv L_\mu - L_\tau$. 
The particle content of our model contains only the SM
particles plus three heavy Majorana neutrinos $N_{1,2,3}$, 
which are singlets of the SM gauge group.  
Under the $U(1)$ symmetry corresponding to $L_\mu - L_\tau$ the three 
flavor eigenstates $\nu_e, \nu_\mu$ and $\nu_\tau$ have the charges 
$0, 1$ and $-1$, respectively. We can assign now to the usual SM 
Higgs doublet $\Phi$ the $L'$ number 0 and 
for the three singlets $N_1, N_2$ and $N_3$ we 
choose $L'=0, 1$ and $-1$, respectively. 
The relevant Lagrangian is then given by a Dirac mass matrix $m_D$ 
connecting the flavor states and the heavy singlets and by a Majorana 
mass matrix $M_R$ for the latter: 
\bea \label{eq:L} 
-{\cal{L}} = \overline{\nu}_\alpha \, (m_D)_{\alpha i} \, N_i + 
\frac{1}{2} N_i^T C^{-1} \, (M_R)_{ij} \, N_j + h.c. \\[0.3cm] 
= \frac{\D \sqrt{2} }{\D v} \, \Phi \, 
\left( 
a \, \overline{\nu}_e  N_1 + b \, \overline{\nu}_\mu  N_2 + 
d  \, \overline{\nu}_\tau N_3 
\right) + 
\frac{M}{2} 
\, \left(  X \, N_1^T C^{-1} N_1 + Y \, N_2^T C^{-1} N_3 \right) + h.c.
\eea
Here, $a,b,d,X,Y$ are real constants of order one, $M$ is the scale of 
the heavy singlets, $v/\sqrt{2}$ is the vacuum expectation value of the 
(lower component) of the Higgs doublet and $C$ is the charge 
conjugation matrix. With our charge assignment given above, the 
charged lepton mass matrix is also real and diagonal. 
After integrating out the heavy 
singlet states, the neutrino mass matrix 
$m_\nu \simeq - m_D^T \, M_R^{-1} \, m_D$ reads 
\be \D
m_\nu \simeq - \frac{v^2}{M} 
\left(
\bad 
\D \frac{a^2}{X} & \D 0 & \D 0 \\[0.3cm]\D
\D \cdot & \D 0 & \D \frac{bd}{Y}  \\[0.3cm]
\D \cdot & \D \cdot & \D 0 
\ea 
\right) ~
\ee
and conserves $L_\mu - L_\tau$. 
The scale of the neutrino masses $m_0 = v^2/M$ can be adjusted by the 
in general unknown scale $M$. Note that to explain the 
global oscillation data one needs
$a^2/X \simeq b d/Y$, which can for instance be achieved by 
$a\simeq b \simeq d$ and $X \simeq Y$. 
A further symmetry is therefore required.  
As discussed before (cf.\ with Table \ref{tab:tab1}), 
this is similar to the case of $L_e$ conservation, 
where the smallness of the determinant of the $\mu\tau$ block 
is required in order to have large solar neutrino 
mixing and to the 
case of $L_e - L_\mu - L_\tau$ where
the two entries $m_{e \mu}$ 
and $m_{e \tau}$ of the mass matrix have to be of the same order
for maximal atmospheric neutrino mixing.  

We assume now that the heavy Majorana mass matrix softly breaks 
$L_\mu - L_\tau$. This can be achieved by adding a democratic perturbation 
to $M_R$. It will then be given by 
\be
M_R = M 
\left( 
\bad 
X + \epsilon & \epsilon & \epsilon \\[0.3cm]
\cdot & \epsilon & Y + \epsilon \\[0.3cm] 
\cdot & \cdot  & \epsilon 
\ea
\right)~,
\ee
where $\epsilon \ll X,Y$. 
It is easy to see that for the light neutrino mass matrix 
the structure of Eq.\ (\ref{eq:mnudem1}) is reproduced, i.e., an 
anarchical perturbation is added to Eq.\ (\ref{eq:mnu0}). 
Assuming $a=b=d$ and $X=Y$ will lead to the particularly simple structure 
of Eq.\ (\ref{eq:mnudem}), where a purely democratic perturbation is added 
to Eq.\ (\ref{eq:mnu0}). 
Of course we can also add a small 
perturbation to $M_R$ which has random order one 
coefficients. Then the structure given in 
Eq.\ (\ref{eq:mnudem1}) will be obtained again. 

We note that the model leading to a mass matrix conserving 
$L_\mu - L_\tau$ presented in \cite{foot} worked with a 
type II see--saw mechanism \cite{typeII} and required a larger particle 
content, namely two Higgs doublets and three triplets. 
A model similar in spirit to ours has been worked out in \cite{grimus}
for the $L_e - L_\mu - L_\tau$ flavor symmetry, using only two 
heavy Majorana neutrinos. 
To successfully reproduce the neutrino oscillation observables, 
the soft breaking terms in $M_R$ had to be of 
the same order as the $L_e - L_\mu - L_\tau$ conserving terms
in \cite{grimus}. 
Our model requires only small breaking terms but requires 
three heavy right--handed neutrinos.

\section{\label{sec:concl}Conclusions}

The flavor symmetry $L_\mu - L_\tau$ for the light neutrino 
mass matrix, on which there 
are very few analyzes, was considered. It predicts one maximal, two zero 
mixing angles and quasi--degenerate neutrinos. 
We showed that demanding the bimaximal mixing scheme and 
quasi--degenerate neutrinos from the most general neutrino mass matrix 
yields a matrix obeying this flavor symmetry. Furthermore, 
a matrix conserving $L_\mu - L_\tau$ is a 
special case of $\mu\tau$ symmetric mass matrices. 
Two simple methods to break $L_\mu - L_\tau$ were presented. 
First, we added a purely democratic term to the neutrino mass matrix 
and applied radiative corrections. Large values of 
$\tan \beta$ were seen to be required to reproduce the correct neutrino 
phenomenology. Atmospheric 
neutrino mixing very close to maximal was predicted. 
Next, we allowed for random coefficients in the 
perturbation matrix added to break the $L_\mu - L_\tau$
symmetry of the 
the zeroth order mass matrix which strictly conserves $L_\mu - L_\tau$. 
Rather large deviations from maximal atmospheric mixing 
could be generated in this scheme.  
For the first scheme of $L_\mu - L_\tau$ symmetry 
breaking, the observables $U_{e3}^2$ and 
$|1/2 - \sin^2\theta_{23}|$ were seen to be proportional to the 
inverse of the fourth power of the common neutrino mass scale, though this 
interesting behavior was spoiled in the second case due to possible 
cancellations caused by the different order one parameters. 
Finally, a simple model based on the see--saw mechanism was 
presented, which reproduced a neutrino mass matrix conserving 
$L_\mu - L_\tau$. 
Soft and small breaking of the flavor symmetry in the heavy 
singlet sector was shown to 
reproduce the two possibilities for symmetry breaking mentioned above.

\vspace{0.5cm}
\begin{center}
{\bf Acknowledgments}
\end{center}
We thank the Scuola Internazionale Superiore di Studi Avanzati, Trieste,
where major part of this work was completed. S.C. thanks INFN for 
financial support during this period. 
This work was supported by the ``Deutsche Forschungsgemeinschaft'' in the 
``Sonderforschungsbereich 375 f\"ur Astroteilchenphysik'' 
and under project number RO-2516/3-1 (W.R.).

\newpage

\end{document}